\documentclass[10pt]{bmc_article}    

\usepackage{cite} % Make references as [1-4], not [1,2,3,4]
\usepackage{url}  % Formatting web addresses  
\usepackage{ifthen}  % Conditional 
\usepackage{multicol}   %Columns
\usepackage{amsthm}
\usepackage{epsfig}
\usepackage{mathrsfs}
\usepackage{appendix}
\usepackage{amsmath,amssymb}
\def \beq {\begin{equation}}
\def \edq {\end{equation}}
\def \bes {\begin{subequations}}
\def \eds {\end{subequations}}
\def \beqn {\begin{equation*}}
\def \edqn {\end{equation*}}

\def \up {\uparrow}
\def \down {\downarrow}

\def \sm {\sigma}

\def \veps {\varepsilon}

\def \calf {{\cal{F}}}

\def\includegraphics{}

\newboolean{publ}

\newenvironment{bmcformat}{\begin{raggedright}\baselineskip20pt\sloppy\setboolean{publ}{false}}{\end{raggedright}\baselineskip20pt\sloppy}

\begin{document}
\begin{bmcformat}


\title{Noise and fluctuation relations of a spin diode}
 
%%%%%%%%%%%%%%%%%%%%%%%%%%%%%%%%%%%%%%%%%%%%%%
%%                                          %%
%% Enter the authors here                   %%
%%                                          %%
%% Ensure \and is entered between all but   %%
%% the last two authors. This will be       %%
%% replaced by a comma in the final article %%
%%                                          %%
%% Ensure there are no trailing spaces at   %% 
%% the ends of the lines                    %%     	
%%                                          %%
%%%%%%%%%%%%%%%%%%%%%%%%%%%%%%%%%%%%%%%%%%%%%%

\author{Jong Soo Lim\correspondingauthor$^{1}$%
       \email{Jong Soo Lim\correspondingauthor - lim.jongsoo@gmail.com}%
      \and
         Rosa L\'opez\correspondingauthor$^{1,2}$%
         \email{Rosa L\'opez\correspondingauthor - rosa.lopez-gonzalo@uib.es}
       and 
         David S\'anchez\correspondingauthor$^{1,2}$%
         \email{David S\'anchez\correspondingauthor - david.sanchez@uib.es}%
      }

%%%%%%%%%%%%%%%%%%%%%%%%%%%%%%%%%%%%%%%%%%%%%%
%%                                          %%
%% Enter the authors' addresses here        %%
%%                                          %%
%%%%%%%%%%%%%%%%%%%%%%%%%%%%%%%%%%%%%%%%%%%%%%

\address{%
    \iid(1)Instituto de F\'{\i}sica Interdisciplinar y Sistemas Complejos IFISC (UIB-CSIC), E-07122 Palma de Mallorca, Spain\\
    \iid(2)Departament de F\'{\i}sica, Universitat de les Illes Balears, E-07122 Palma de Mallorca, Spain
}%

\maketitle

%%%%%%%%%%%%%%%%%%%%%%%%%%%%%%%%%%%%%%%%%%%%%%
%%                                          %%
%% The Abstract begins here                 %%
%%                                          %%
%% The Section headings here are those for  %%
%% a Research article submitted to a        %%
%% BMC-Series journal.                      %%  
%%                                          %%
%% If your article is not of this type,     %%
%% then refer to the Instructions for       %%
%% authors on http://www.biomedcentral.com  %%
%% and change the section headings          %%
%% accordingly.                             %%   
%%                                          %%
%%%%%%%%%%%%%%%%%%%%%%%%%%%%%%%%%%%%%%%%%%%%%%

\begin{abstract}
We consider fluctuation relations between the transport coefficients
of a spintronic system where   magnetic interactions play a crucial role.
We investigate a
prototypical spintronic device---a spin-diode---which consists of an 
interacting resonant level coupled to two ferromagnetic electrodes. 
We thereby obtain the cumulant generating function for the 
spin transport in the sequential tunnelling regime. 
We demonstrate the fulfilment of the nonlinear fluctuation relations when
up and down spin currents are correlated in the presence
of both spin-flip processes and external magnetic fields. 
\end{abstract}

\ifthenelse{\boolean{publ}}{\begin{multicols}{2}}{}

%%%%%%%%%%%%%%%%
%% Background %%
%%
\section*{Background}
Nonequilibrium fluctuation relations overcome the limitations of linear response theory
and yield a complete set of relations that connect different transport coefficients out of equilibrium
using higher-order response functions \cite{RevModPhys.81.1665,PhysRevB.72.235328,PhysRevLett.101.046802,PhysRevB.78.115429,PhysRevLett.101.136805,PhysRevB.79.045305,PhysRevLett.104.076801}.
Even in the presence of symmetry breaking fields,
it is possible to derive nonlinear fluctuation relations from the microreversibility principle applied to the scattering matrix at equilibrium \cite{PhysRevLett.101.136805}. 
A possible source of time-reversal symmetry breaking are magnetized leads.
Then, it is necessary to include in the general formulation the spin degree of freedom,
which is an essential ingredient in spintronic applications \cite{Igor:04} such as
spin-filters \cite{Recher:00} and spin-diodes \cite{PhysRevLett.92.206801,PhysRevB.75.165303,1674-4926-31-6-062002,%
PhysRevB.70.115315,PhysRevB.62.1186,PhysRevB.84.115304,PhysRevB.74.075328,PhysRevB.77.075305}.

We recently proved nonequilibrium fluctuation relations valid for spintronic systems \cite{rosi},
fully taking into account spin-polarized leads, magnetic fields and spin-flip processes.
Here, we investigate a spin diode system and explicitly
demonstrate that the spintronic fluctuation relations are satisfied. Furthermore, we calculate
the spin noise (correlations of the spin polarized currents) and discuss its main properties.

\section{Theoretical model}

Consider a quantum dot coupled via tunnel barriers to two ferromagnetic leads $\alpha=L,R$, as shown in Fig. 1 (a).
The leads have spin-dependent density of states $\rho_{\alpha\up}(\omega) \ne \rho_{\alpha \down}(\omega)$ [flat density of states  are depicted Fig. 1(a)]. 
For convenience, we introduce the leads' spin polarization parameter as $p_{\alpha} = (\rho_{\alpha \up} - \rho_{\alpha\down})/(\rho_{\alpha \up} + \rho_{\alpha\down})$.  
In the limit of $\Delta \veps \gtrsim k_BT,|eV|$ ($\Delta\veps$ is the dot level spacing, $k_B$ the Boltzmann constant and $T$ the temperature) effectively only a single energy level $\veps_{\sm}$
($\sm=\uparrow,\downarrow$) in the dot contributes to transport and can be occupied by $0$, $1$, or $2$ electron charges.
In the presence of an external magnetic field $B$, the  Zeeman splitting  is $\veps_{\up} - \veps_{\down} = g\mu_B B$ ($g$ is the Land\'{e} factor and $\mu_B=q\hbar/2 m$ is the Bohr magneton with $q$ the electron charge). 
Tunneling between lead $\alpha$ and the dot yields a level broadening given by $\Gamma_{\alpha \sm}(\omega) = \pi \rho_{\alpha \sm}|V_{\alpha}|^2$ ($V_{\alpha}$ is the lead-dot tunneling amplitude).
Notice that the level width is then spin-dependent due to the spin asymmetry of the density of states: $\Gamma_{\alpha\sm} = (\Gamma/2)(1 + s p_{\alpha})$ with $\Gamma = \Gamma_L = \Gamma_R$ and $s=+$($-$) for $\uparrow$($\downarrow$).

\begin{figure}
\centerline{
\epsfig{file=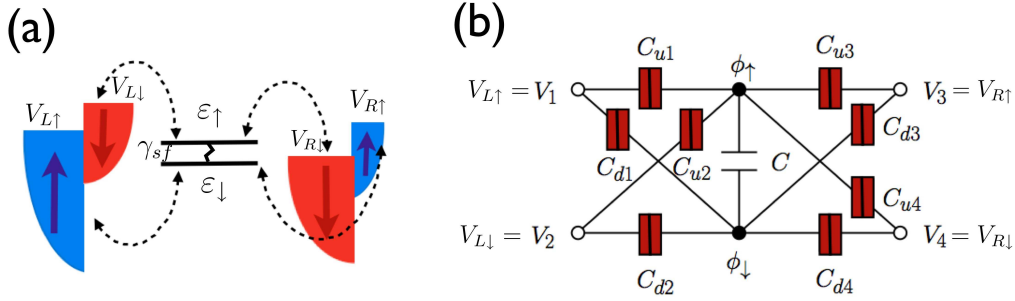,angle=0,width=0.85\textwidth,clip}
}
\caption{(a)  Sketch of the spin diode system. The dot level is attached to two ferromagnetic contacts. 
$V_{L\sm}$, and $V_{R\sm}$ indicate the spin dependent bias voltages applied to the left ($L$), and ($R$) right contacts, respectively. 
The dot level is spin split by a magnetic field $B$: $\veps_\uparrow\neq\veps_\downarrow$. 
Both spin dependent energy levels are connected by spin-flip processes with a rate given by $\gamma_{sf}$. 
(b) Electrostatic model: $\phi_{\up}$, and $\phi_{\down}$ are the dot internal potentials calculated using capacitance couplings [$C_{ui}$, $C_{di}$ ($i=1\cdots 4$), $C$] within an electrostatic model. }
\end{figure}

In the limit of weak dot-lead coupling, $\Gamma \ll k_BT $, tunneling occurs sequentially and transport is thus dominated by
first-order tunnelling processes. The dynamics of the system is governed by the time evolution of the occupation probabilities calculated from the master equation $dP/dt = \mathcal{W} P$, 
with $P \equiv \{P_0,P_{\up},P_{\down},P_2\}$ denoting the probabilities associated to states with $0$ electrons on the dot, $1$ electron with spin $\up$ or $\down$ and $2$ electrons. 
We also take into account spin-flip relaxation mechanisms possibly present in our system due to magnetic interactions with a spin fluctuating environment  
(e.g., hyperfine coupling with nuclear spins) or spin-orbit interactions in the dot: $\gamma_{sf}^{\sigma\bar\sigma} = \gamma_{sf} 
\exp\left[(\veps_{\sigma}-\veps_{\bar\sigma})/(2k_B T)\right]$. To study the full counting statistics of a  spin diode we consider the generalized
rate transition matrix $\mathcal{W}(\chi)$, with $\chi=\{\chi_{L\uparrow},\chi_{L\downarrow},\chi_{R\uparrow},\chi_{R\downarrow}\}$ the counting fields:
\begin{eqnarray}\label{system}
\mathcal{W}(\chi)=
\begin{pmatrix}
-\sum_{\alpha,\sm} \Gamma_{\alpha\sm}^+ & \sum_{\alpha} \Gamma_{\alpha\up}^-e^{i\chi_{\alpha\up}} & \sum_{\alpha} \Gamma_{\alpha\down}^-e^{i\chi_{\alpha\down}} & 0 \\
\sum_{\alpha} \Gamma_{\alpha\up}^+e^{-i\chi_{\alpha\up}} & -\sum_{\alpha}\Gamma_{\alpha\up}^- - \sum_{\alpha}\tilde{\Gamma}_{\alpha\down}^+ - \gamma_{sf}^{\up\down} & \gamma_{sf}^{\down\up} & \sum_{\alpha} \tilde\Gamma_{\alpha\down}^-e^{i\chi_{\alpha\down}} \\
\sum_{\alpha} \Gamma_{\alpha\down}^+e^{-i\chi_{\alpha\down}} & \gamma_{sf}^{\up\down} & -\sum_{\alpha}\Gamma_{\alpha\down}^- - \sum_{\alpha}\tilde\Gamma_{\alpha\up}^+ - \gamma_{sf}^{\down\up} & \sum_{\alpha} \tilde\Gamma_{\alpha\up}^-e^{i\chi_{\alpha\up}} \\
0 & \sum_{\alpha} \tilde\Gamma_{\alpha\down}^+e^{-i\chi_{\alpha\down}} & \sum_{\alpha} \tilde\Gamma_{\alpha\up}^+e^{-i\chi_{\alpha\up}} & -\sum_{\alpha,\sm} \tilde\Gamma_{\alpha\sm}^-
\end{pmatrix}\,,
\end{eqnarray}
%where $\{\chi\} \equiv \{\chi_1,\chi_2,\chi_3,\chi_4\}$ are the counting fields and fulfil $\chi_1=\chi_3$ and $\chi_2=\chi_4$ and 
where $\Gamma_{\alpha\sm}^{\pm} = \Gamma_{\alpha\sm} f^{\pm}(\mu_{0\sm}-eV_{\alpha\sm})$,
$\tilde{\Gamma}_{\alpha\sm}^{\pm} = \Gamma_{\alpha\sm} f^{\pm}(\mu_{1\sm}-eV_{\alpha\sm})$,
and $f^{\pm}(\veps) = 1/[\exp(\pm\veps/k_BT)+1]$.
Here, $V_{\alpha\sigma}$ is a spin-dependent voltage bias and
$\mu_{i\sm}$ the dot electrochemical potential to be determined from the electrostatic model.
$i=0, 1$ an index that takes into account the charge state of the dot.
Then, the cumulant generating function in the long time limit is given by $\calf(\chi;t) = \lambda_0(\chi) t$
where $\lambda_0(\chi)$ denotes the minimum eigenvalue of $\mathcal{W}(\chi)$ that develops adiabatically from $0$ with $\chi$. From the generating function all
transport cumulants are obtained \cite{rosi}.

We consider a gauge-invariant
electrostatic model that treats interactions within a mean-field approach \cite{PhysRevB.72.201308}.
For the geometry sketched in Fig. 1(b) we employ the discrete Poison equations for the charges $Q_{\up}$ and $Q_{\down}$:
$Q_{\up} = C_{u1}(\phi_{\up} - V_{L\up}) + C_{u2}(\phi_{\up} - V_{L\down}) + C_{u3}(\phi_{\up} - V_{R\up}) + C_{u4}(\phi_{\up} - V_{R\down}) + C(\phi_{\up} - \phi_{\down})$ 
and $Q_{\down} = C_{d1}(\phi_{\down} - V_{L\up}) + C_{d2}(\phi_{\down} - V_{L\down}) + C_{d3}(\phi_{\down} - V_{R\up}) + C_{d4}(\phi_{\down} - V_{R\down}) + C(\phi_{\down} - \phi_{\up})$, 
where $C_{\ell i}$ represent capacitance couplings.
We then find the potential energies for both spin orientations, $U_{\sigma}(N_{\sigma},N_{\bar\sigma}) = \int_0^{qN_{\sigma}} dQ_{\sigma}~ \phi_{\sigma}(Q_{\bar\sigma},Q_{\sigma})$, $N_{\sigma}$ being the excess electrons in the dot. 
For an empty dot, i.e., $N_{\up} = N_{\down} = 0$, its electrochemical potential for the spin $\up$ or $\down$ level can be written as $\mu_{0\sigma} = \veps_{\sigma} + U_{\sigma}(1,0) - U_{\sigma}(0,0)$. 
This is the energy required to add one electron into the spin $\up$ or $\down$ level when both spin levels are empty.
%where $\mu_{\sigma\alpha} = E_F + qV_{\sigma\alpha}$ is the electrochemical potential of lead $\alpha$ and $E_F$ the  common Fermi energy.
Importantly, our results are gauge invariant since they depend on potential differences ($V_{\alpha\sm} - V_{\alpha'\sm'}$) only.
When the dot is charged, then  $N_{\up} = 1$ or $N_{\down} = 1$ and we find
$\mu_{1\sm} = \mu_{0\sm} + 2q^2/\widetilde{C}$, with $\widetilde{C} = K/C$ and $K=\sum_i C_{ui} \sum_j C_{dj} + C\sum_{\ell=u/d} \sum_i C_{\ell i}$.

\section{Results}
\subsection{Nonlinear fluctuation relations}

We denote with $\alpha,\beta,\gamma$ both the lead index and the spin channel. Thus, $\alpha=1$ corresponds to lead $L$ and spin $\uparrow$,
$\alpha=2$ corresponds to lead $L$ and spin $\downarrow$, etc.; see Fig. 1.
Let $I_\alpha$ be the current operator which accounts for the spin flow in a given terminal.
Then, the $I$--$V$ characteristics reads, up to second order in voltage,
\begin{eqnarray}
\langle I _{\alpha}\rangle=\sum_{\beta} G_{\alpha,\beta} V_\beta +\frac{1}{2}\sum_{\beta\gamma} G_{\alpha,\beta\gamma}V_\beta V_\gamma\,,
\end{eqnarray}
where $\langle\cdots\rangle$ is a quantum mechanical average.
Current-current correlations (noise)  between fluctuations $\Delta I=I-\langle I\rangle$
are calculated up to first order in voltage:
\beq
S_{\alpha\beta}\equiv\langle \Delta I_\alpha \Delta I_\beta\rangle =S_{\alpha\beta}^{(0)}+\sum_{\gamma} S_{\alpha\beta,\gamma} V_\gamma\,.
\edq
Small fluctuations around equilibrium and their responses are related through the fluctuation-dissipation theorem. 
In particular, the Kubo formula for the electrical transport relates the linear conductance $G_{\alpha,\beta}$ (electrical response)
to the equilibrium noise $S_{\alpha\beta}^{(0)}$ (equilibrium current fluctuation). 
Relations among the transport coefficients that appear in a nonlinear voltage expansion of the high order current cumulants have been recently obtained for spintronic systems \cite{rosi}. 
Thus, in the weakly nonlinear transport regime we find that the equilibrium third current cumulant,
$\mathcal{C}_{\alpha\beta\gamma}^{(0)}$, is related to the second order non-linear conductance,
$G_{\alpha,\beta\gamma}$,
and the  noise susceptibilities, $S_{\alpha\beta,\gamma}$, by means of a fluctuation relation,
\begin{equation}\label{nonlineal}
\mathcal{C}_{\alpha\beta\gamma}^{(0)} = k_BT\left(S_{\alpha\beta,\gamma} + S_{\alpha\gamma,\beta} + S_{\beta\gamma,\alpha}\right) -(k_BT)^2\left(G_{\alpha,\beta\gamma} + G_{\beta,\alpha\gamma} + G_{\gamma,\alpha\beta} \right)\,.
\end{equation}
%In the following, we give analytical expressions for the spin noises in the spin diode system and their corresponding Fano factors. %Additionally, we explicitly check the fulfilment of the nonlinear fluctuation relations in Eq. (\ref{nonlineal}).

We analyze a quantum dot attached to both a ferromagnetic lead with polarization $p_L=p$ and a normal lead with polarization $p_R=0$. We take into account the presence of spin-flip processes described by $\gamma_{sf}$.
In Fig. 2 we explicitly check the fulfilment of Eq.\ \eqref{nonlineal} for different values of the lead polarization in the general case 
of a spin-dependent bias configuration: $V_{L\uparrow}=V_1$, $V_{L\downarrow}=V_2$,  $V_{R\uparrow}=V_3$, $V_{R\downarrow}=V_4$. 
When the dot is subjected to an externally applied magnetic field, one must consider the antisymmetrized version of Eq.\ (\ref{nonlineal}) using $A_-=A(B)-A(-B)$, where $A$ can be $G$, $S$ or higher order correlation functions
($\mathcal{C}_{-}^{(0)}=0$ for an energy independent scattering matrix as in our system). Importantly, the checked relations involve terms of current cross correlations at different spin channels. The occurrence of nonvanishing cross correlations appears when
spin-flip processes correlate the spin channels.  Remarkably, only when these cross correlations are not zero the nonlinear relations are nontrivially satisfied.

\begin{figure}
\centering
\centering
\epsfig{file=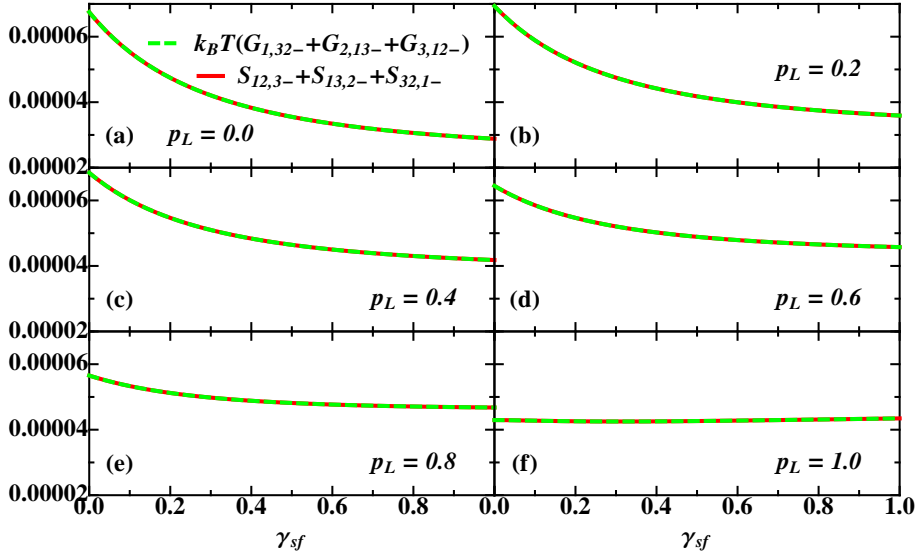,angle=0,width=0.85\textwidth,clip} 
\caption{Verification of spintronic fluctuation relations, Eq. (\ref{nonlineal}). 
Parameters: $\Gamma_0 = 1$, $q^2/C_0 = 40\Gamma_0$ ($C_{ui}=C_{di}=C_0$), $C=\infty$, $\veps_d = 0$, $p_L \ne 0$, $p_R = 0$, $k_BT = 5\Gamma_0$, and $g\mu_B B=0.1\Gamma_0$.}
\end{figure}

\subsection{Spin noise}

We now discuss the analytical expressions for the spin noises of our spin diode.
We consider that the system is biased with a source-drain voltage $V_{SD}=V_1-V_3$ with $V_{1}=V_{2}$ and $V_{3}=V_{4}$.
For definiteness, we take the limit $C\to\infty$ (double occupation is forbidden)
and zero magnetic field ($\veps_{\uparrow}=\veps_\downarrow$). Then, we are able to obtain an analytical expression
for the cross correlations between $\uparrow$ and $\downarrow$ currents in the left terminal:
\begin{align}
S_{L\up L\down} &= 
\begin{cases}
-\frac{2}{27}(1-p^2)\Gamma_0 & \text{for $qV_{SD} > +2|\veps_{eff}|$} \\
-\frac{2(1-p^2)\left[(1-3p^2) + 6\gamma_{sf}/\Gamma_0 + 12(\gamma_{sf}/\Gamma_0)^2 + 8(\gamma_{sf}/\Gamma_0)^3\right]}{(3-p^2)^3 + 6\gamma_{sf}/\Gamma_0}\Gamma_0 & \text{for $qV_{SD} < -2|\veps_{eff}|$}
\end{cases}
\end{align}
where $\veps_{eff}=\veps+e^2/2C_{\Sigma}$ with $C_{\Sigma} = \sum_{\ell,i}C_{\ell i}$.
When the level lies inside the transport window the 
cross-correlations are suppressed as $p$ increases independently of $\gamma_{sf}$.
Moreover, $S_{L\up L\down}$ is always negative due to the antibunching behaviour of fermions \cite{PhysRevB.46.12485}.
The shot noise diagonal in the spin indices is given by
\begin{align}
S_{L\up L\up} &= 
\begin{cases}
\frac{1}{27}\left(7+(5-2p)p\right)\Gamma_0 
& \text{for $qV_{SD} > +2|\veps_{eff}|$}
\\
\frac{(1+p)\left[(1-p)+2\gamma_{sf}/\Gamma_0\right]\left[(7+6p-2p^3+p^4)-4(2p^2+p-7)\gamma_{sf}/\Gamma_0+4(7-2p)(\gamma_{sf}/\Gamma_0)^2\right]}{\left[(3-p^2)+6\gamma_{sf}/\Gamma_0\right]^3}\Gamma_0
& \text{for $qV_{SD} < -2|\veps_{eff}|$}
\end{cases}
\end{align}
with an associated Fano factor  $F_{L\up L\up}=S_{L\up L\up}/I_{L\up}$,
\beq
F_{L\up L\up} =
\begin{cases}
1 - \frac{2}{9}(1+p) & \text{for $qV_{SD} > +2|\veps_{eff}|$} \\
1 + \frac{2(1+p)\left[(4p-p^2-1) + 2(p-2)\gamma_{sf}/\Gamma_0 - 4(\gamma_{sf}/\Gamma_0)^2\right]}{\left[(3-p^2) + 6\gamma_{sf}/\Gamma_0\right]^2} & \text{for $qV_{SD} < -2|\veps_{eff}|$}
\end{cases}
\edq
Notably, the Fano factor is always sub-Poissonian whenever $\veps_{eff}$ lies inside the transport window.
This is due to correlations induced by Coulomb interactions \cite{PhysRevLett.92.106601}.

\section{Conclusions}
Nonequilibrium fluctuation relations nicely connect nonlinear conductances with noise susceptibilities.
We have derived spintronic fluctuation relations for a prototypical spintronic system: a
spin diode consisting of a quantum dot attached to two ferromagnetic contacts.
We have additionally investigated the fulfilment of such relations when both
spin-flip processes inside the dot and an external magnetic field are present in the sample. We have also
inferred exact, analytical expressions for the spin noise current correlations and the Fano factor.
Further extensions of our work might consider noncollinear magnetizations and energy dependent
tunneling rates.

%%%%%%%%%%%%%%%%%%%%%%%%%%%
\section*{Acknowledgements}
  \ifthenelse{\boolean{publ}}{\small}{}
This work was supported by MINECO Grants No.
FIS2011-2352 and CSD2007-00042 (CPAN), 
CAIB and FEDER.

%%  Bmc_article.bst  will be used to                       %%
%%  create a .BBL file for submission, which includes      %%
%%  XML structured for BMC.                                %%

{\ifthenelse{\boolean{publ}}{\footnotesize}{\small}
 \bibliographystyle{bmc_article}  % Style BST file
  \bibliography{papertnt2_new} }     % Bibliography file (usually '*.bib' ) 

%%%%%%%%%%%

\ifthenelse{\boolean{publ}}{\end{multicols}}{}

%%%%%%%%%%%%%%%%%%%%%%%%%%%%%%%%%%%
%%                               %%
%% Figures                       %%
%%                               %%
%% NB: this is for captions and  %%
%% Titles. All graphics must be  %%
%% submitted separately and NOT  %%
%% included in the Tex document  %%
%%                               %%
%%%%%%%%%%%%%%%%%%%%%%%%%%%%%%%%%%%

%%
%% Do not use \listoffigures as most will included as separate files

%\bibliography{bmc_article.bib}
\end{bmcformat}
\end{document}